\documentclass[aps,prd,twocolumn,groupedaddress,floatfix]{revtex4}
\input epsf

\usepackage{amssymb}

\begin{document}

\preprint{{\bf MADPH-03-1351}}

\title{Universal light quark mass dependence and\\ heavy-light meson
spectroscopy} 

\author{Theodore J. Allen} \affiliation{Department of Physics, Hobart
\& William Smith Colleges, \\ Geneva, NY 14456 USA}

\author{Todd Coleman} 

\affiliation{Joint Science Department, Claremont McKenna, Pitzer, and
Scripps Colleges, \\ 925 N. Mills Ave., Claremont, CA 91711 USA }

\author{M.G. Olsson}

\affiliation{Department of Physics, University of Wisconsin, \\ 1150
University Avenue, Madison, WI 53706 USA }

\author{Sini\v{s}a Veseli}

\affiliation{Fermi National Accelerator Laboratory, \\ P.O. Box 500,
Batavia, IL 60510 USA}

\begin{abstract}

Clean predictions are presented for all the spin-averaged heavy-light meson
spectroscopies.  A new symmetry is identified wherein the energy
eigenstates have a universal dependence on both the light and heavy
quark masses.  This universality is used in an efficient analysis of
these mesons within the QCD string/flux tube picture.  Unique
predictions for all the $D,\ D_s,\ B$, and $B_s $ type mesons in terms
of just four measured quantities.

\end{abstract}

\maketitle

\section{Introduction}
One of the most promising and least developed areas of hadron
spectroscopy is the excited heavy-light (HL) meson.  Although at
present only a few states of each flavor have been observed future
discoveries at $B$ factories, CLEOc, HERA, and hadron colliders will
surely change this situation.  In addition to the ordinary $q\bar Q$
states we should observe hybrid and possibly multi-quark confined
mesons.  It is important therefore to reliably predict where the
standard HL mesons lie and to explore the close relationship between
the $D, D_s, B$ and $B_s$ families of HL mesons.

A striking observed fact for HL systems is that hyperfine splittings
are independent of light quark flavor.  For example \cite{ref:PDG}
\begin{eqnarray}
                 && D^*_s - D_s\ \simeq\ D^* - D\ \simeq\ 142\rm\
                 MeV\,, \\ && B^*_s - B_s\ \simeq\ B^* - B\ \simeq\
                 \phantom046\rm\ MeV \,.
\end{eqnarray}
This apparent lack of light quark mass dependence in these differences is
certainly not that expected in the popular Breit-Fermi type
semi-relativistic interaction which is proportional to the inverse product
of the quark masses ($1/m_Q m $).  In the following we show that this is an
example of a larger ``universal light quark mass dependence'' (UMD)
ultimately a consequence of relativistic kinematics.  We note in passing
that the ratio of the above hyperfine differences does however reflect the
inverse ratio of the heavy quark masses.

We start by proposing and supporting the concept that all HL energy
eigenstates have the same light quark mass dependence and hence all
differences containing the same light flavor are independent of light quark
mass. We take this as an organizing principle to analyze the various HL
systems.  In particular we find a functional relation between strange and
non-strange light quark masses, $m_s$ and $m_{u,d}$, and we predict the spectra
for the $D,\ D_s,\ B$, and $B_s$ systems.  In Sec.\ II we establish the UMD
principle first from experiment, then from a model calculation. We finally
exhibit the UMD within a simple potential model.  In Sec.\ III we use UMD
to determine the parameters of the relativistic flux tube, a simple but
fundamental model.  These parameters are the Coulomb constant and the heavy
quark masses $m_c$ and $m_b$.  Another application of UMD is given in Sec.\
IV where from the measured difference $B_s - B$, a relationship between the
constituent quark masses $m_s$ and $m_{u,d}$ is established.  This relationship
is also shown to follow from relativistic kinematics alone.  In Sec.\ V we
use our results to predict a range of radially and orbitally excited HL
mesons.

\section{Universal Light Quark Mass Dependence}\label{sec:two}

The HL meson mass $M$, in the heavy quark limit, can be defined in
terms of the excitation energy $E$ and the heavy quark mass $m_Q$ as
\begin{equation}\label{eq:old3}
                         M=m_Q + E     \,.
\end{equation}
As we will demonstrate, the meson mass $M$ has universal mass dependence on
both the heavy quark mass $m_Q$ and the light quark mass $m$.  Up to
$1/m_Q$ corrections, we may expand $E$ as
\begin{equation}\label{eq:old4}
             E_{n,\ell}= E_{n,\ell}(0) + \beta m^2 +\dots   \,,
\end{equation}
where the coefficient $\beta$ is independent of both the radial number $n$
and the angular quantum number $\ell$.  We expect the expansion to have
only even powers of $m$ since $m$ only appears quadratically in our
model Hamiltonians.  The energy differences between different HL
excitations is then
\begin{equation}\label{eq:old5}
           E_2-E_1 = E_2(0) - E_1(0) + \beta (m_2^2 - m_1^2)+\dots .
\end{equation}
The excitation energy differences of HL mesons with the same light flavor
are independent of the quark mass.  We offer three types of evidence for
this (UMD) universality.

\subsection{Experimental  Data}

We select any convenient $P$-wave and $S$-wave $D$ type meson
difference \cite{ref:PDG}.  For example,
\begin{equation}\label{eq:old6}
                D_1(2422\pm2{\rm\ MeV}) - D(1864\pm0.5{\rm\ MeV}) =
                558\pm2\rm\ MeV \,.
\end{equation}  
We now compare this with the corresponding $D_s$ difference,
\begin{equation}\label{eq:old7}
           D_{s1}(2535\pm0.5{\rm\ MeV}) - D_s(1969\pm1.4{\rm\ MeV}) =
           566\pm1.5\textrm{ MeV} \,.
\end{equation}
If UMD is valid, the two differences should be identical.
Experimentally they differ by $8\pm3$ MeV which is an accuracy of
better than 2\%.

Other differences involving $D^*$ and $D_2$ give similar results but
with slightly larger errors.

\subsection{A dynamical model: the relativistic flux tube}

The Relativistic Flux Tube (RFT) (or QCD string) model with spinless
quarks has been solved numerically for about a decade
\cite{ref:asymmetric}.  For a rigorous derivation and experimental
motivation see \cite{ref:Reduction}.  We will not discuss the details
of this model here except to emphasize that it is a very realistic
model incorporating many of the features of QCD.  In addition to the
string confinement (with static tension $a$) we add a short range
interaction $U(r)=-k/r$.  In the heavy quark limit the heavy quark
mass appears additively as in Eq.~(\ref{eq:old3}) with no $1/m_Q$
corrections.  The light quark constituent mass is not well known and
we will only assume that it is in the range $0<m< 600$ MeV.  The
parameters that appear in the RFT model are the string tension $a$,
the Coulomb constant $k$, and the masses of the heavy quark, $m_Q$, and
the light quark, $m$.  Since the quarks are spinless in this model we
will always compare our model predictions to the spin-averaged data.

In this sub-section we are exploring the properties of the RFT model
and not comparing to experimental data so the exact values of the
parameters are not important.  We use $a=0.18\rm\ GeV^2$ and $k=0.5$
which are in fact reasonable values, as we discuss in the next section.
In Fig.~1 we plot the lowest $S$ and $P$ wave eigenvalues of the
excitation energy $E$ as a function of the light quark mass $m$.  The
important thing to notice is that the difference $1P-1S$ is quite
constant.  This is exactly what is expected under UMD as in
Eqs.~(\ref{eq:old4}) or~(\ref{eq:old5}).

\begin{figure}[htb]
\epsfxsize=\linewidth
\epsfbox{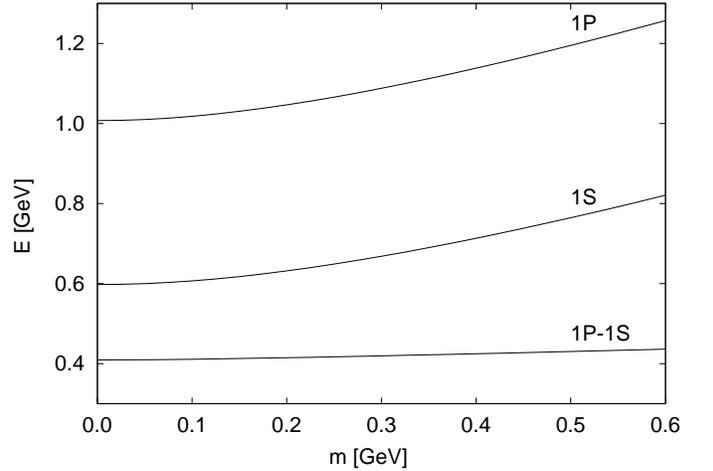}
\caption{Light quark mass dependence of the lowest $S$ and $P$-wave
heavy-light excitation energies as predicted by the relativistic flux tube
model.  We note that the $1P-1S$ mass difference is nearly independent of
light quark mass.}\label{fig:1}
\end{figure}

\subsection{A simple analytic result}

We show here that UMD is fundamentally a result of relativistic
kinematics.  Let us consider a simple time-component vector potential
model with relativistic kinematics,
\begin{eqnarray}\label{eq:old8}
                        H\psi &=& E\psi ,    \\    
\label{eq:old9}         H&=& \sqrt{\textbf{p}^2+m^2} +U(r) ,
\end{eqnarray}
where
\begin{equation}\label{eq:old10}
                        \textbf{p}^2=p_r^2+L^2/r^2   \,.
\end{equation}
An expression of UMD is
\begin{equation}\label{eq:old11}
                        \partial^2E/\partial L^2 \partial m^2  =0     \,.      
\end{equation}
We can demonstrate this to leading order with the Feynman-Hellmann
theorem \cite{ref:Feynman}
\begin{equation}\label{eq:old12}
                        \partial E/\partial \lambda = \left< \partial
                        H/\partial \lambda \right> \,,
\end{equation}
where $\lambda$ is a system parameter.  The desired quantity
(\ref{eq:old11}) is then given by
\begin{equation}\label{eq:old13}
                         \partial^2E/\partial L^2 \partial m^2 =
\left< \partial ^2H/\partial L^2 \partial m^2 \right> \,.
\end{equation}
Expanding about $L^2 = m^2 =0$ we find from (\ref{eq:old9}) that
\begin{equation}\label{eq:old14}
                H= p_r + (L^2/r^2 + m^2 )/(2 p_r) +\dots         
\end{equation}
which then yields to leading order
\begin{equation}\label{eq:old15}
                         \partial^2E/\partial L^2 \partial m^2  =0 .
\end{equation}
The above demonstration of light quark universality also holds for
mesons with two light quark mesons.

The next term in the expansion (\ref{eq:old14}) is a cross term
proportional to $L^2\,m^2$, which yields a non-vanishing second derivative
(\ref{eq:old15}) that violates UMD.  A similar violation using the RFT
model can be seen in Fig.~1 and amounts to about 10 MeV for $m$ increasing
from 300 MeV to 500 MeV.  This accounts for the small observed violation of
$8\pm 3$ MeV noted in Eqs.~(\ref{eq:old6}) and (\ref{eq:old7}).

\section{The RFT parameters  \lowercase{$a$, $k$, $m_c$, $m_b$}}

As we have noted the parameters entering the RFT model are the string
tension $a$, the Coulomb constant $k$, and the two heavy quark masses $m_c$
and $m_b$. The predictions for excited states will be nearly independent of
the light quark mass value but sensitive to the difference between $m_s$
and $m_{u,d}$.

\subsection{String tension}

The universal Regge slope for both mesons and baryons is \cite{ref:Barger},
\begin{equation}\label{eq:old16}
                         \alpha' =0.88\rm\  GeV^2                        
\end{equation}
For a relativistically rotating QCD string, the Nambu-Goto QCD string
\cite{ref:Nambu} and the RFT model predict the Regge slope to be
\begin{equation}\label{eq:old17}
                      \alpha' = 1/2\pi a \,,
\end{equation}
which yields the string tension,
\begin{equation}\label{eq:old18}
                           a=0.18\rm\ GeV^2   \,.
\end{equation}
This value is quite consistent with that obtained from an analysis of
heavy onia data alone \cite{ref:Jacobs} and we will assume it in our
subsequent work.

\subsection{Coulomb constant}

Without any assumptions about the quark masses, we can find the Coulomb
constant using UMD as an organizing principle.  The idea is to compare the
model predictions with an experimental $1P-1S$ HL mass difference.  To do
this we must know the masses of a pair of spin averaged states.
Fortunately, we now have a complete set of states for the $D_s$ mesons due
to recent $B$ factory measurements \cite{ref:CLEO} and older data
\cite{ref:PDG}.  We find the spin averaged (weighted by angular momentum
multiplicity) states to be
\begin{eqnarray}\label{eq:old19}
          D_{s,1S} &=& \frac34 D^*_s + \frac14 D_s = 2076\pm1\rm\ MeV \ , \\ 
 D_{s,1P} &=& \frac1{12} D_{s,0+}
    +\frac14 (D_{s,1+}^{1/2}+D_{s,1+}^{3/2})+\frac5{12} D_{s,2+} \nonumber\\
\label{eq:old21}              &=& 2515\pm3\ \textrm{MeV} \ .
\end{eqnarray}
We need the difference
\begin{eqnarray}
\label{eq:old22}    D_{s,1P} - D_{s,1S} =439\pm4\rm\ MeV     .
\end{eqnarray}
In Fig.~2 we show the RFT prediction for this difference as a
function of the Coulomb constant $k$ and we see that the correct value
is,
\begin{equation}\label{eq:old23}
                           k\simeq 0.52\,.
\end{equation}

\begin{figure}[htb]
\epsfxsize=\linewidth
\epsfbox{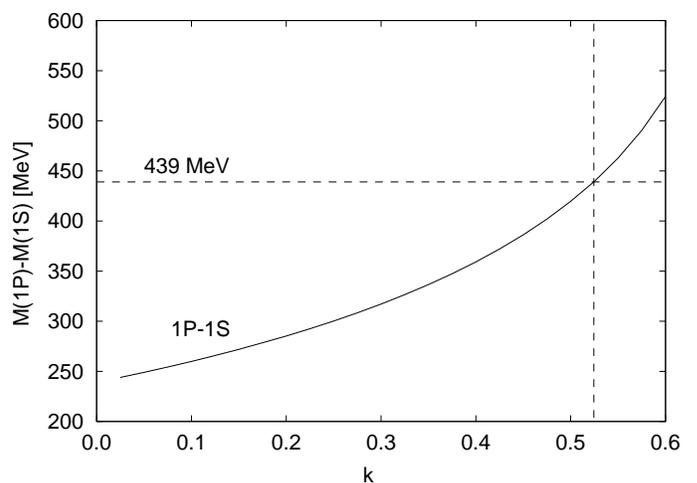}
\caption{The difference of the lowest $S$ and $P$ wave heavy-light meson
masses in the RFT model as a function of the Coulomb constant $k$.  The
horizontal line is the experimental value (\protect\ref{eq:old22})
determined from the $D_s$ states.}\label{fig:2}
\end{figure}

\subsection{Heavy quark masses}

The heavy quark masses do depend on the choice of light quark mass.
In order to agree with the observed $1S$ state we must adjust $m_Q$ as
$m$ is varied.  The results for the charm and bottom quarks are shown in
Figs.~3 and 4.

\begin{figure}[htb]
\epsfxsize=\linewidth
\epsfbox{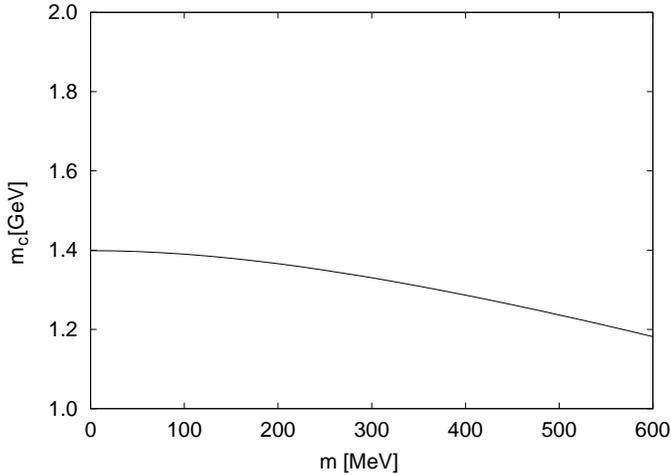}
\caption{The  $c$ quark mass required to yield the observed spin averaged
$D_{1S}$  meson mass for a range of choices of light quark mass.}\label{fig:3}
\end{figure}

\begin{figure}[htb]
\epsfxsize=\linewidth
\epsfbox{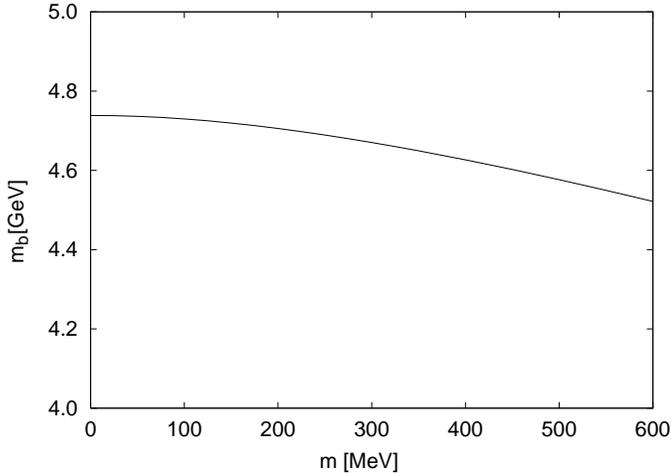}
\caption{The $b$  quark mass required to yield the observed spin averaged
 $B_{1S}$ meson mass for a range of choices of light quark mass.}\label{fig:4}
\end{figure}

\section{Relation between the light quark masses}

For a given heavy quark, say the $b$, the two ground state mesons $B_s$
and $B$ are expected to differ in mass as given in Eq.~(\ref{eq:old5})
since $\beta$ is not zero.  Experimentally this difference is
\cite{ref:PDG}
\begin{equation}\label{eq:old24}
                     B_s - B = 91\pm1\rm\  MeV\,.
\end{equation}
This observation implies a functional relation between the strange and
non-strange quark masses.  Using the values for $a$ and $k$ in Eqs.\
(\ref{eq:old18}) and (\ref{eq:old23}), we exhibit this relation for the RFT
model in Fig.~5.  We might note that the corresponding charm difference is
about 10 MeV larger and reflects a larger heavy quark kinetic energy
(\textit{i.e.\/}, a $1/m_Q$ correction).  Finally, we might comment that
this relation between quark masses again arises primarily from relativistic
kinematics.  The relation
\begin{equation}\label{eq:old25}
                \sqrt{p_0^2+m_s^2} - \sqrt{p_0^2+m_{u,d}^2} = 91 \rm\ MeV          
\end{equation}
follows from the simple Hamiltonian (\ref{eq:old9}).  With the choice
$p_0^2=0.4$ GeV$^2$ as the average square momentum, the implicit
relationship of Eq.~(\ref{eq:old25}) parallels that of the more realistic
RFT nicely and is depicted on Fig.~5 by the dashed curve.  We also include
the solutions for the light quark constituent masses obtained from analyses
of hyperon magnetic moments.  We note that these ``hyperon'' values of the
light quark masses fall close to our curve.

\begin{figure}[thb]
\epsfxsize=\linewidth
\epsfbox{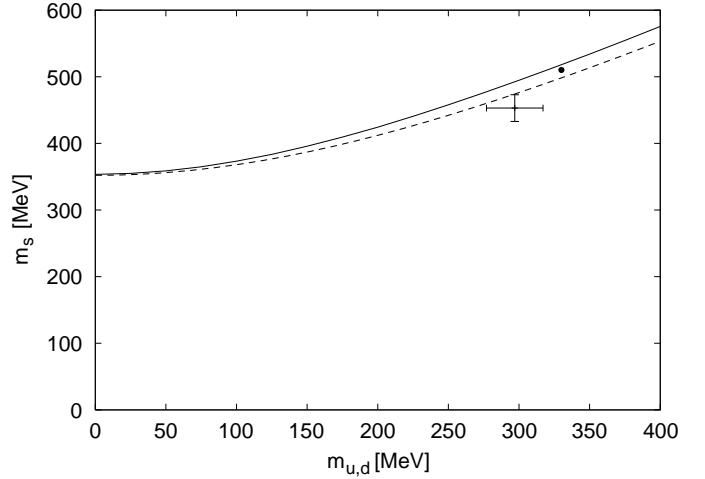}
\caption{The relation between the strange and the non-strange light quark
masses required to maintain the relation (\protect\ref{eq:old24}).  The
``data'' points corresponds to the well known and successful hyperon
magnetic moment model where the quarks have Dirac moments inversely
proportional to the constituent quark masses \cite{ref:PDG} (dot) and
\cite{ref:Franklin} (error bar).  The solid curve is the prediction of the
relativistic flux-tube (RFT) model.  The dashed curve is computed using the
pure relativistic kinematics of Eq.\ (\protect\ref{eq:old25}) with
$p_0^2=0.4$ GeV.}\label{fig:5}
\end{figure}

\section{Predictions for the heavy-light spectra}

Now that the parameters of the RFT model have been fixed, we can make a
range of predictions.  We note that we have required only the $S$-wave
spin averaged ground states for the charm and bottom states, and one
spin averaged $P$-wave multiplet (in our case the $D_s$).  The
predictions are then unique and independent of specific choices of the
light quark masses.  In Figs.~6 to 9 we present our predictions for
the $D, D_s, B$, and $B_s$ flavor families.  In each case we predict
up to five radial and five angular states.  As we expect from UMD, the
predictions are nearly unique.  If the light quark mass is varied over
a 200 MeV range, the predictions for the excited states vary by less
than 10 MeV, which would be difficult to see on the figures.  In Table
1 we provide the numerical predictions for the spin-averaged states
assuming $m_{u,d}=300$ MeV and $m_s=500$ MeV.

\begin{figure}[htb]
\epsfxsize=\linewidth
\epsfbox{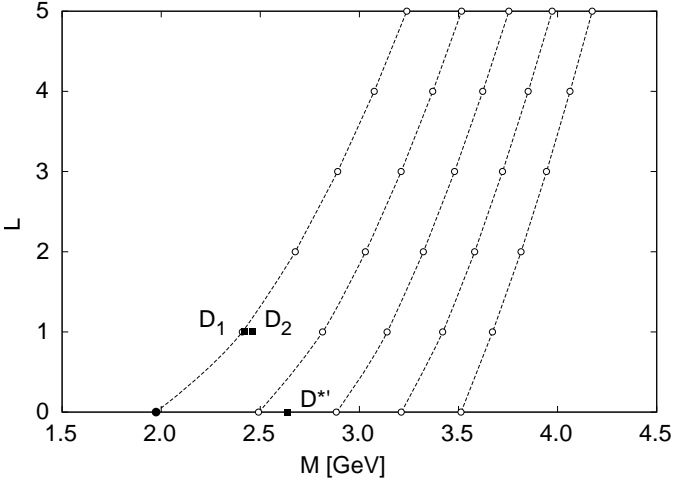}
\caption{The $D (cu,cd)$ spectroscopy.  The large solid dot is input
data and the hollow dots are predicted states. The predicted states
are displayed numerically in Table 1. The boxes represent measured
(spin) states not used in the calculation.  The $D_1 (2420)$ and $D_2
(2460)$ are well known but the $D^{*\prime}(2637)$ \protect\cite{ref:delphi}
should be verified.}\label{fig:6}
\end{figure}

\begin{figure}[htb]
\epsfxsize=\linewidth
\epsfbox{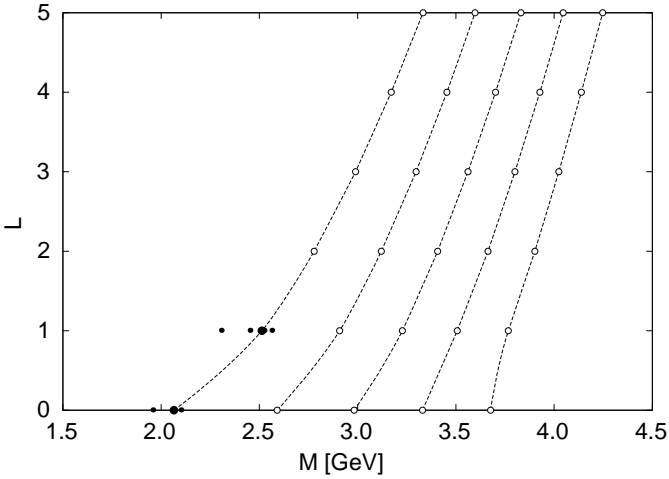}
\caption{The $D_s (cs)$ spectroscopy.  The large solid dots are input
data and the hollow dots are predicted states.  The predicted states
are displayed numerically in Table 1. To illustrate the spin
dependence we show the 1S and 1P spin states as small dots.} \label{fig:7}
\end{figure}

\begin{figure}[htb]
\epsfxsize=\linewidth
\epsfbox{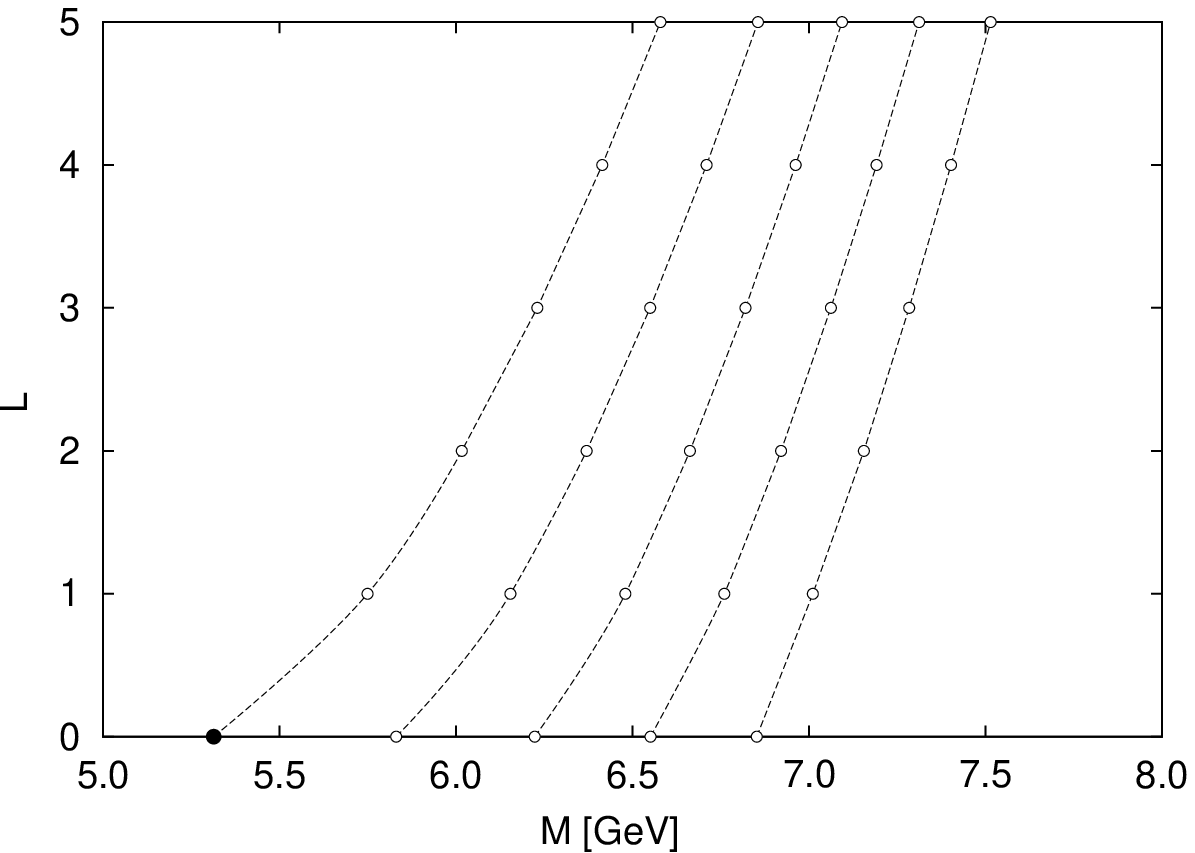}
\caption{The $B (bu,bd)$ spectroscopy.  The solid dot is input data and
the hollow dots are predicted states. The predicted states are
displayed numerically in Table 1. }\label{fig:8}
\end{figure}

\begin{figure}[htb]
\epsfxsize=\linewidth
\epsfbox{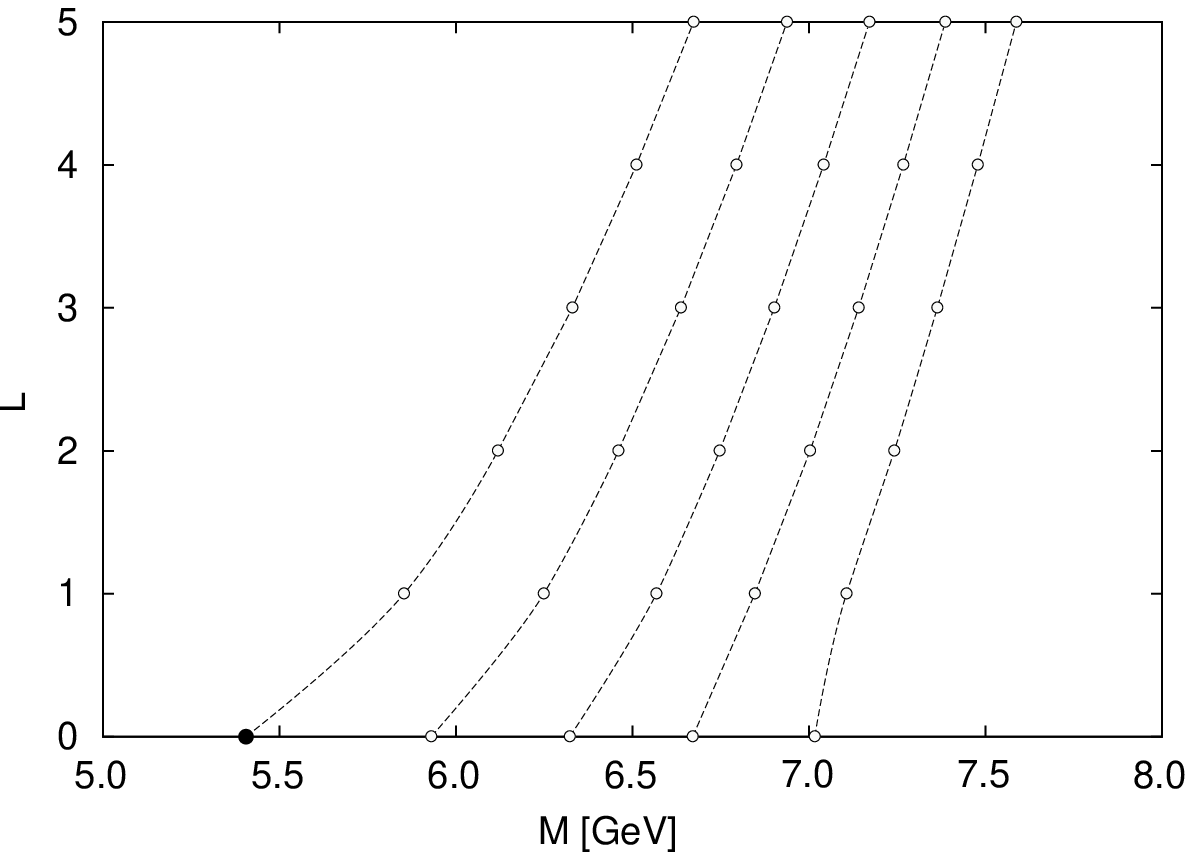}
\caption{The $B_s (bs)$ spectroscopy.  The solid dot is input data
and the hollow dots are predicted states. The predicted states are
displayed numerically in Table 1.}\label{fig:9}
\end{figure}

\begin{table}
\tabcolsep1em
\caption{Predicted heavy-light meson states in GeV. The flux tube
parameters are $a=0.18$ GeV$^{-2}$, $k=0.524$, $m_{u,d}=300$ MeV, $m_s=495$
MeV, $m_c=1330$ MeV, $m_b=4670$ MeV.  These states are illustrated in
Figs.\ 6--9.  Varying the light quark masses over a range of 200 MeV
changes our predictions by less than 10 MeV.}
\medskip
\begin{tabular}{|c|cccc|}
\hline
  & \multicolumn{4}{c|}{$n$}\\
  & 1& 2& 3& 4\\
\hline
$\ell$&    \multicolumn{4}{c|}{$D$ states}\\
0 &        1.974&  2.491&  2.883&  3.211\\
1 &        2.409&  2.814&  3.140&  3.420\\
2 &        2.677&  3.030&  3.323&  3.581\\
3 &        2.891&  3.210&  3.480&  3.722\\
\hline
  &    \multicolumn{4}{c|}{$D_s$  states}\\
0 &       2.065&  2.590&  2.982&  3.331\\
1 &       2.513&  2.909&  3.228&  3.507\\
2 &       2.779&  3.120&  3.407&  3.663\\
3 &       2.990&  3.297&  3.562&  3.801\\
\hline
  &      \multicolumn{4}{c|}{$B$ states}\\
0&        5.314& 5.831& 6.223& 6.551\\
1&        5.749& 6.154& 6.480& 6.760\\
2&        6.017& 6.370& 6.663& 6.921\\
3&        6.231& 6.550& 6.820& 7.062\\
\hline
&        \multicolumn{4}{c|}{$B_s$  states}\\
0&        5.405& 5.930& 6.322& 6.671\\
1&        5.853& 6.249& 6.568& 6.847\\
2&        6.119& 6.460& 6.747& 7.003\\
3&        6.330& 6.637& 6.902& 7.141\\
 \hline
\end{tabular}
\end{table}

\section{Conclusions}

We have approached the subject of heavy-light meson spectroscopy by
introducing a new principle which we call ``universal light quark mass
dependence'' (UMD).  The idea is that the energies of all orbitally and
radially excited states vary in the same way as the light quark mass is
varied.  This proposal is supported in Sec.\ \ref{sec:two} by experimental
evidence, numerical calculations using a realistic theoretical model, and
finally by analytic demonstration using a simple but relativistic potential
model.  This universality observation makes the analysis of heavy-light
mesons transparent and considerably simpler.  We further note that the
measured $B_s - B$ difference implies a functional relation between the
strange and non-strange light quark masses which is consistent with the
well known quark model of the hyperon magnetic moments
\cite{ref:PDG,ref:Franklin}.  From only the $S$-wave states and one
spin-averaged $P$-wave state, the $D_s$, we can reliably predict the $D$,
$D_s$, $B$, and $B_s$ excited spectrum.  It should be noted that our
predictions are for the spin-averaged states and that we assumed that the
heavy-light assumption is valid.  There are some small discrepancies in the
data when thought of in the heavy-light limit.  For example $D_s - D$ is
about 10 percent larger (10 MeV) than the corresponding difference $B_s
-B$. This is probably due to $1/m_Q$ corrections to heavy quark symmetry.
Another topic for future investigation is the breakdown of the heavy-light
approximation for highly excited states.


\section*{Acknowledgment}
This work was supported in part by the US Department of Energy under
Contract No.~DE-FG02-95ER40896.


\vfill

\begin{thebibliography}{99}

\bibitem{ref:PDG} K. Hagiwara, {\it et al.,\/} Phys.\ Rev.\ D {\bf 66}, 010001
(2002).

\bibitem{ref:asymmetric} D. La Course and M. G. Olsson, Phys.~Rev.~D
{\bf 39}, 2751 (1989); M. G. Olsson and S. Veseli, Phys.~Rev.~D {\bf
51}, 3578 (1995); C.~Semay and B.~Silvestre-Brac,
Phys.\ Rev.\ D {\bf 52}, 6553 (1995).

\bibitem{ref:Reduction} T. J. Allen and M. G. Olsson, Phys.\ Rev.\ D {\bf 68},
032002 (2003) [hep-ph/0306128].


\bibitem{ref:Feynman} H. Hellmann, {\it Einf\"uhrung in die
Quantenchemie}, (Deuticke Verlag, Leipzig, 1937); R.P. Feynman,
Phys.~Rev.\ {\bf 56}, 340 (1939).

\bibitem{ref:Barger} V. Barger and D. Cline, {\it Phenomenological
Theories of High Energy Scattering: An Experimental Evaluation},
(W. A. Benjamin, New York, 1969).

\bibitem{ref:Nambu} Y. Nambu, {\it ``Quark model and the factorization of
the Veneziano Amplitude,''} in {\it Symmetries and quark models,\/}
edited by R.~Chand, (Gordon and Breach, New York, 1970); T. Goto, {
Prog.~Theor.~Phys.\/} {\bf 46}, 1560 (1971); L. Susskind, {Nuovo
Cimento\/} {\bf 69A}, 457 (1970); A. M. Polyakov, { Phys.\ Lett.\ B\/}
{\bf 103}, 207 (1981); B. M. Barbashov and V. V. Nesterenko, {\it
Introduction to the Relativistic String Theory\/}, (World Scientific,
Singapore, 1990).

\bibitem{ref:Jacobs} Steve Jacobs, M.G. Olsson, and Casimir Suchyta III,
Phys.~Rev.~D {\bf 33}, 3338 (1986).

\bibitem{ref:CLEO} The narrow $P$-wave $D_s(2460)$ and $D_s(2317)$
have recently been observed at the CLEO (D. Besson, {\it et al.,\/} Phys.\
Rev.\  D {\bf 68}, 032002 (2003) [hep-ex/0305100]), BABAR, and
Belle detectors.

\bibitem{ref:Franklin} Jerrold Franklin, Phys.~Rev.~D {\bf 66} 033010 (2002) .

\bibitem{ref:delphi} P. Abreu, {\it et al.\/} (DELPHI Collaboration), 
Phys.~Lett.~B {\bf 426}, 231 (1998).

\end{thebibliography}
\end{document}